# Evidence of Nodal Superconductivity in $Na_{0.35}CoO_2 \cdot 1.3H_2O$: A Specific Heat Study

H. D. Yang[1], J.-Y. Lin[2], C. P. Sun[1], Y. C. Kang[1], C. L. Huang[1], K. Takada[3], T. Sasaki[3], H. Sakurai[4] and E. Takayama-Muromachi[4]

[1]*Department of Physics, National Sun Yat Sen University, Kaohsiung 804, Taiwan, R.O.C.*
[2]*Institute of Physics, National Chiao Tung University, Hsinchu 300, Taiwan, R.O.C.*
[3]*Soft Chemistry Research Group, Advanced Materials Laboratory, National Institute for Materials Science, 1-1 Namiki, Tsukuba, Ibaraki 305-0044, JAPAN*
[4]*Superconducting Materials Center, National Institute for Materials Science 1-1 Namiki, Tsukuba, Ibaraki 305-0044, Japan*

Comprehensive low-temperature specific heat data $C(T,H)$ of $Na_{0.35}CoO_2$ $1.3H_2O$ with temperature $T$ down to 0.6 K and the magnetic field $H$ up to 8 T are presented. In the normal state, the values of $\gamma_n$=13.94 mJ/mol K$^2$, and Debye temperature $\Theta_D$=362 K are reported. At zero field, a very sharp superconducting anomaly was observed at $T_c$=4.5 K with $\Delta C/\gamma_{ns}T_c$=1.45 if the specific heat jump is normalized to the superconducting volume fraction, which is estimated to be 47.4 % based on the consideration of entropy balance at $T_c$ for the second-order superconducting phase transition. In the superconducting state, the electronic contribution $C_{es}$ at $H$=0 can be well described by the model of the line nodal order parameter. In low $H$, $\delta\gamma(H) \propto H^{1/2}$ which is also a manifestation of the line nodes. The behaviors of both $T_c(H)$ and $\gamma(H)$ suggest the anisotropy of $H_{c2}$ or possible crossovers or transitions occurring in the mixed state.

$Na_xCoO_2 \cdot yH_2O$ [1] is one of the most interesting superconductors since the high-temperature superconductors (HTSC's) [2] and $MgB_2$ [3] were discovered. But unlike $MgB_2$, which superconducting mechanism was largely understood within one year of its discovery, some of the fundamental questions about $Na_xCoO_2 \cdot yH_2O$ remain open at this moment in defiance of intensive theoretical and experimental efforts. The parent compound $Na_xCoO_2$ is known to be a strongly correlated electron system. By the intercalation of $H_2O$ molecules between $CoO_2$ planes, quasi two-dimensional superconductivity is induced in $CoO_2$ planes similar to that in $CuO_2$ planes of cuprates. On the other hand, with the triangular $CoO_2$ planes rather than the nearly square $CuO_2$ planes, there possibly exists new superconductivity. The theoretical studies thus follow at least two approaches. Some propose that $Na_xCoO_2 \cdot yH_2O$ is a resonating valence bond superconductor [4-8], closely related to HTSC's. Others suggest new mechanisms like charge fluctuations, which could make $Na_xCoO_2 \cdot yH_2O$ a novel superconductor [9]. In principle, experimental studies of $Na_xCoO_2 \cdot yH_2O$ could help distinguish some theoretical models from the others. However, the experimental studies so far have shown some contradictory results of pairing symmetry and its spin state even by the same technique (for example, NQR and NMR [10-12]).

The specific heat ($C$) technique can probe the bulk properties of the samples and has been proven to be a powerful tool to investigate the pairing state of novel superconductors such as high-$T_c$ cuprates [13-16], $MgB_2$ [17,18], and $MgCNi_3$ [19]. $C(T,H)$ also provides the information about the quasiparticle excitation associated with the mixed state in magnetic fields. Although specific heat measurements of $Na_xCoO_2 \cdot yH_2O$ were reported in several works [20-24], few of them presented data lower than 2 K and with the magnetic field ($H$) dependence. However, the low-$T$ ($T<2$ K) data and those in $H$ are supposed to be crucial to the elucidation of superconductivity in such a superconductor with $T_c \sim 4.5$ K. To shed light on this important issue, we have measured $C(T,H)$ of $Na_xCoO_2 \cdot yH_2O$ with temperature down to 0.6 K and in magnetic fields up to 8 T on several samples which were made in different batches. Typical results and analyses for one of the samples are presented and clearly show that the order parameter in $Na_xCoO_2 \cdot yH_2O$ is unconventional.

Polycrystalline $Na_xCoO_2 \cdot yH_2O$ powder was prepared and characterized as described in [1]. The composition was determined to be $x=0.35$ and $y=1.3$. Thermodynamic $T_c$ determined from $C(T)$ is 4.5 K for the present sample (see below). $C(T)$ was measured using a $^3$He thermal relaxation calorimeter from 0.6 K to 10 K in magnetic fields $H$ up to 8 T. A detailed description of the measurements can be found in Ref. [18]. Prior to the measurement, the powder was kept in the environment of almost 100 % relative humidity with saturated NaCl solution. This treatment could be very crucial to preserve the water content and consequently the superconducting volume fraction. It was then cold pressed by applying a pressure of about $1.6 \times 10^4$ N cm$^{-2}$ into pellets with ~ $1.5 \times 1.5 \times 0.3$ mm$^3$ in size and ~ 2mg in mass for $C$ measurements. One sample was measured two days after the first run of the specific heat measurements. Both runs rendered identical $C(T)$ within the resolution limit of the apparatus indicating the stability of the samples at temperature of liquid helium.

$C(T)$ of $Na_{0.35}CoO_2 \cdot 1.3H_2O$ down to 0.6 K with $H=0$ to 8 T is shown in Fig. 1 as $C/T$ vs. $T^2$. A pronounced superconducting anomaly was observed at $T_c \sim 4.5$ K at $H=0$ indicating that the bulk superconductivity in the present sample is similar to that reported in Refs. [20-24], and persists with $H$ at least up to 6 T. To further analyze the data, the first step is to quantify the normal state specific heat $C_n(T)$. $C_n$ can be written as $C_n(T)=\gamma_n T+C_{lattice}$, where $C_{lattice}=ST^3+DT^5$ represents the phonon contribution. Naively, one may try to obtain $C_n$ by fitting the data above $T_c$. However, a more elaborate analysis is to take the entropy balance of the second order phase transition into consideration. This further analysis results in $\gamma_n=13.94\pm0.21$ mJ/mol K$^2$, $S=0.295\pm0.007$ mJ/mol K$^4$ (corresponding to the Debye temperature $\Theta_D=362$ K) and $D=(1.6\pm0.6)\times10^{-4}$ mJ/mol K$^6$. The resultant $C_n$ is shown as the solid curve in Fig. 1, and the entropy balance is achieved as depicted in the inset of Fig. 2 by the integration of $\delta C/T \equiv C(H=0)/T - C_n/T$ with respect to $T$ from $T=0$ to $T_c$. It is interesting to note that $C_n(T)$ determined by this way is very close to the data of $H=8$ T as shown in Fig. 1.

It is noted that $C/T$ does not extrapolate to zero as $T$ approaches zero at $H=0$ suggesting that the superconducting volume is less than 100% (Fig. 1). However, the peak is as sharp as that observed in many other well identified superconductors [17-19, 25]. Therefore, the existence of a well separated superconducting portion in the sample rather than a broad spread in $T_c$ can be taken as a plausible assumption. The extrapolation of the solid line (the line nodal superconductivity model, discussed later) in Fig. 2 leads to $\delta C(T=0)/T=-6.61$ mJ/mole K$^2$. Considering only the superconducting fraction in the sample, the corresponding value of $\gamma_{ns}$ should be appropriately taken as 6.61 mJ/mol K$^2$, which is associated with the carriers participating in the superconducting transition, rather than 13.94 mJ/mol K$^2$ which includes additional contribution from nonsuperconducting fraction. In this context, the volume fraction of the superconducting portion can be estimated by $(-\delta C(T=0)/T)/\gamma_n = \gamma_{ns}/\gamma_n = 6.61/13.94 = 47.4\%$. This superconducting volume fraction is comparable to that of the best samples in the early era of HTSC's, and is larger than many of the reported values in Refs. [20,24] presumably due to the improved treatment of the sample quality and sample handling technique. The *normalized* dimensionless specific-heat jump at $T_c$ is then $\Delta C/\gamma_{ns}T_c \approx 43.1/(6.61 \times 4.5)=1.45$. This value of 1.45 is close to 1.43 expected for isotropic $s$-wave and is larger than ~1 of line nodal superconductivity, both in weak limit, respectively [27].

Fruitful information of the superconductivity in Na$_{0.35}$CoO$_2$ 1.3H$_2$O can be further deduced from $\delta C(T)/T$ shown in Fig. 2. The thin solid, dashed, and thick solid lines are the fits according to the observed specific-heat jump by the model of the isotropic $s$-wave with $2\Delta/kT_c=3.5$ (weak coupling) and $2\Delta/kT_c=3.8$ (moderate coupling), and the line nodal order parameters [26] with $2\Delta/kT_c=5.0$ (strong coupling), respectively. The data from $T_c$ down to $T$=0.6 K are well described by the model of the line nodal order parameter. On the other hand, the thin solid and dashed lines of $s$-wave pairing deviate significantly from the data, especially at low temperatures. This deviation is due to the power law behavior of the data in $T$ in contrast to the exponential $T$ dependence in the $s$-wave scenario. This result of line nodal superconductivity in $Na_{0.35}CoO_2$ $1.3H_2O$ is also consistent with the recent muon spin relaxation measurements [27]. However, the analysis does not support the scenario of the order parameter with point nodes as suggested in Ref. [20]. At low temperatures, $\delta C(T)/T$ is approximately linear with respect to $T$ shown in Fig. 2. This behavior strongly suggests an $rT^2$ term ($r$ is a constant) in the superconducting electronic specific heat $C_{es}$ as $T \ll T_c$, which is a characteristic of the line nodal order parameter as seen in HTSC's [13-17]. Though there might be sources of uncertainty in the $rT^2$ term from nonsuperconducting fraction, the observed $r$=1.76 mJ/mol K$^3$ in the present sample can be compared with $r$=1.02 mJ/mol K$^3$ in other sample we measured with 26.6% superconducting volume fraction. The scaling of the value of $r$ with the superconducting volume fraction strongly suggests that the $rT^2$ term in $C_{es}$ is an intrinsic property of superconductivity in $Na_{0.35}CoO_2$ $1.3H_2O$. Furthermore, it is of interest to compare the observed $r$ with the estimated coefficient $r \approx \chi_{ns}/T_c$ within nodal superconductivity scenario [15]. The observed $r$=1.76 mJ/mol K$^3$ is in good agreement with the estimated $r \approx 6.61/4.5$=1.47 mJ/mol K$^3$. Similar agreement was also observed in other line nodal superconductors such as $La_{1.78}Sr_{0.22}CuO_4$ [14] and $Sr_2RuO_4$ [26]. Actually, the $rT^2$ term appears in all the superconducting $Na_xCoO_2 \cdot yH_2O$ samples we have measured.

In Fig. 1, no significant magnetic contribution such as the paramagnetic centers was observed in this sample in contrast to that observed in other samples we measured. This allows one to reliably analyze in field data. Figure 3 shows $\delta C(T,H)/T$ of $Na_{0.35}CoO_2$ $1.3H_2O$ in magnetic fields up to 8 T. $H$ gradually suppresses superconductivity with increasing quasiparticle contribution in $C_{es}$ in the mixed state. The entropy balance for the data at each field was also checked and less than 10 % imbalance was observed for all fields. To further quantify the discussion, $T_c(H)$ of $Na_{0.35}CoO_2$ 1.3$H_2O$ is shown in Fig. 4. In low $H$, there is a change of the slope in the $T_c$-$H$ curve at $H$~0.5 T. This slope change was actually observed in several polycrystalline samples from different sources by either $C$ or $M$ measurements [28,29], and appears to be genuine. The two dashed lines in Fig. 4 give qualitative descriptions to the empirical fit of the small and large slopes near $T_c$ with $H_{c2}(0)$ ~ 4 and 20 T, respectively. These two values of $H_{c2}$ from different slopes are consistent with those of $H_{c2}//c$ and $H_{c2}//ab$ from experiments on single crystals [22,30,31]. In higher fields $H$>2 T, the faster $T_c$ suppression than estimation from the large slope is also consistent with the single crystal experiments, and suggests a Pauli paramagnetic limit $H_p$≈8.3 T. Therefore, a possible source of the slope change could be the anisotropy of $H_{c2}$ along different crystalline directions. Another proposed scenario is an $H$-induced phase transition, probably from the singlet to triplet pairing [29]. How this scenario reconciles with the high $H$ results of $T_c(H)$ and $\lambda(H)$ deserves further investigation.

Figure 5 shows the $H^{1/2}$ dependence of $\lambda(H)$ obtained from the linear extrapolation of the data from $T$≤1.5 K down to $T$=0 in Fig. 1. A rapid increase of $\lambda(H)$ in low $H$ is followed by a very slow increase in 0.5 T<$H$<2T. Furthermore, $d\lambda(H)/dH$ becomes large again for $H$>2 T. The $H$ dependence of $\lambda(H)$ in Fig. 5 actually reveals the corresponding mixed-state behavior of $T_c(H)$ in Fig. 4. In principle, the quasiparticle contribution to $C_{es}$ should increase correspondingly with $T_c$ suppression in $H$. Therefore, the consistency between the results in Figs. 4 and 5 convincingly suggests the crossovers or transitions in the mixed state. This complexity in the mixed state could partially resolve the discrepancies in different NQR and NMR experiments [10-12]. More quantitatively, it can be clearly seen in Fig. 5 that the low $H$ data follow $\delta\gamma(H)\equiv\lambda(H)-\lambda(0)\propto H^{1/2}$ until H>0.5 T. This $H^{1/2}$ dependence is a manifestation of nodal line order parameter and has been observed in HTSC's [13-16] and $Sr_2RuO_4$ [25,26]. The dashed line is a fit of data for $H$≤0.5T to $\gamma(H)=\gamma(0)+AH^{1/2}$. This fit leads to $A=3.32$ mJ/mol K$^2$ T$^{1/2}$. This experimental value is in good agreement with the theoretical estimation on $A$ of the line nodal superconductivity, where $A \approx \gamma_{ns}/H_{c2}^{1/2}=6.61/4^{1/2}=3.3$ mJ/mol K$^2$ T$^{1/2}$ [14,32] and $H_{c2}//c=4$ T is taken from Fig. 4. It is interesting to note that this fitted $\gamma(H)$ projects to the normal state $\gamma_n=13.94$ mJ/mol K$^2$ at $H=3.95$ T, not far from the previous estimated $H_{c2}//c=4$ T (see Fig. 5).

To conclude, the comprehensive specific heat studies on high quality Na$_{0.35}$CoO$_2$ 1.3H$_2$O polycrystalline sample have established several fundamental properties of the superconducting Na$_{0.35}$CoO$_2$ 1.3H$_2$O. Both $\delta C/T$ at $H=0$ and $\gamma(H)$ in low magnetic fields provide convincing evidence of nodal lines in the superconducting order parameter. The in-field data further suggest anisotropy in $H_{c2}$ or possible crossovers or transitions in the mixed state. Elucidation of these properties would certainly benefit future theoretical and experimental research on this interesting superconductor.

We are grateful to C. Y. Mou, T. K. Lee, and B. Rosenstein for indispensable discussion. This work was supported by National Science Council, Taiwan, Republic of China under contract Nos. NSC93-2112-M-110-001 and NSC93-2112-M-009-015.

# Captions

Fig. 1. $C/T$ vs. $T^2$ for $Na_{0.35}CoO_2 \cdot 1.3H_2O$ both at zero field and in magnetic fields $H$. The normal state specific heat $C_n(T)$ is denoted as the solid line.

Fig. 2. $\delta C/T \equiv C(H=0)/T - C_n/T$ vs. $T$ for $Na_{0.35}CoO_2 \cdot 1.3H_2O$. The thin solid, dashed and thick solid lines are the fits according to the weak-coupling isotropic $s$-wave, moderate-coupling isotropic $s$-wave, and the line nodal order parameters, respectively. Inset: The entropy difference $\Delta S$ is calculated by integrating $\delta C(T)/T$ with respect to $T$ according to the data above 0.6 K and the solid line below 0.6 K.

Fig. 3. $\delta C(H)/T \equiv C(H)/T - C_n/T$ vs. $T$ for $Na_{0.35}CoO_2 \cdot 1.3H_2O$ in magnetic fields $H$ up to 8 T.

Fig. 4. $T_c(H)$ is determined thermodynamically from $C(T,H)$. The dashed lines are the empirical estimates of the small and large slopes near $T_c$ (see text). Inset shows the example how $T_c(H)$ is determined by the local entropy balance for $H=1$ T data.

Fig. 5. $\chi(H)$ vs. $H^{1/2}$. The dashed line is the linear fit representing $\delta\chi(H) \propto H^{1/2}$ for the data with $H \leq 0.5$ T. The horizontal dot line denotes the normal state $\chi_n$.

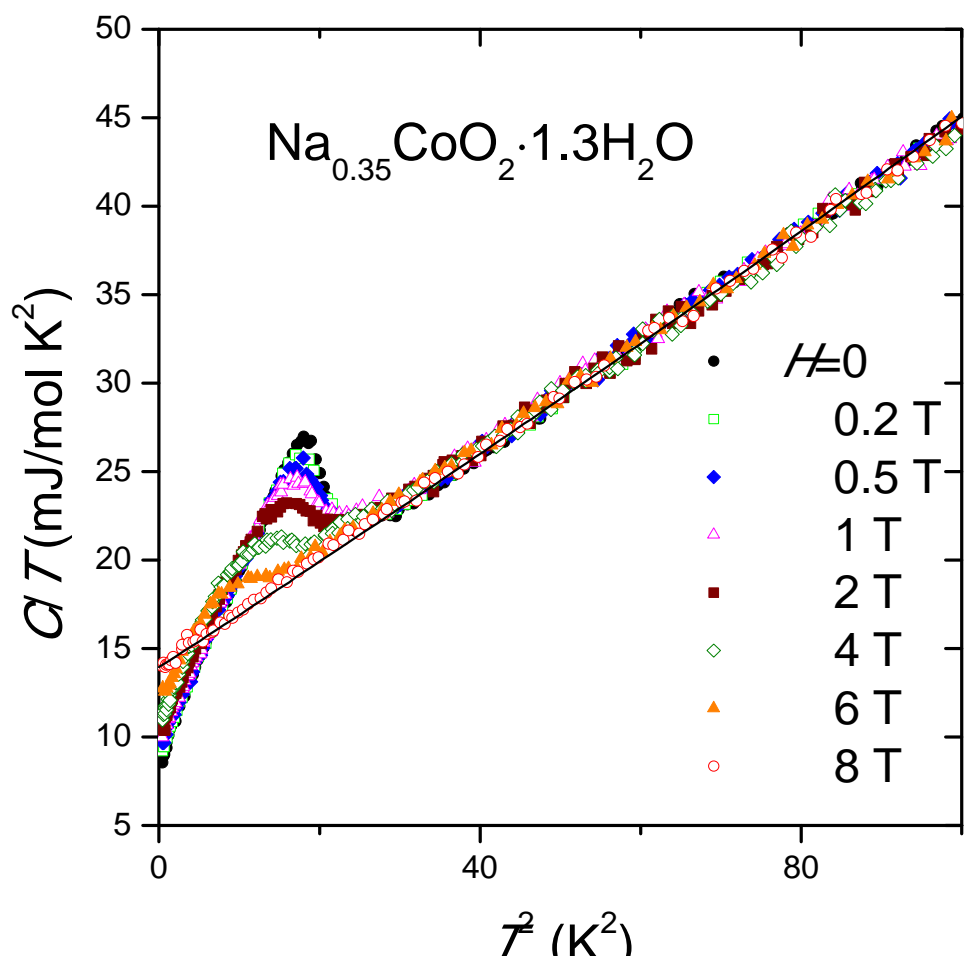

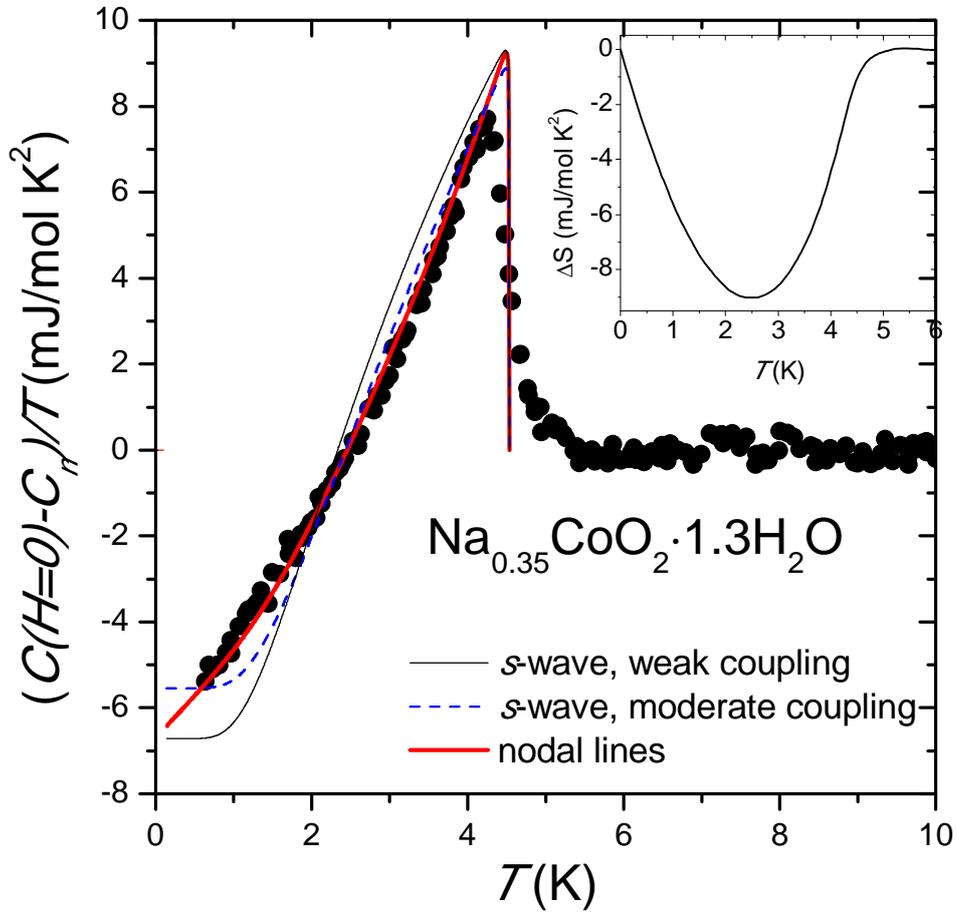

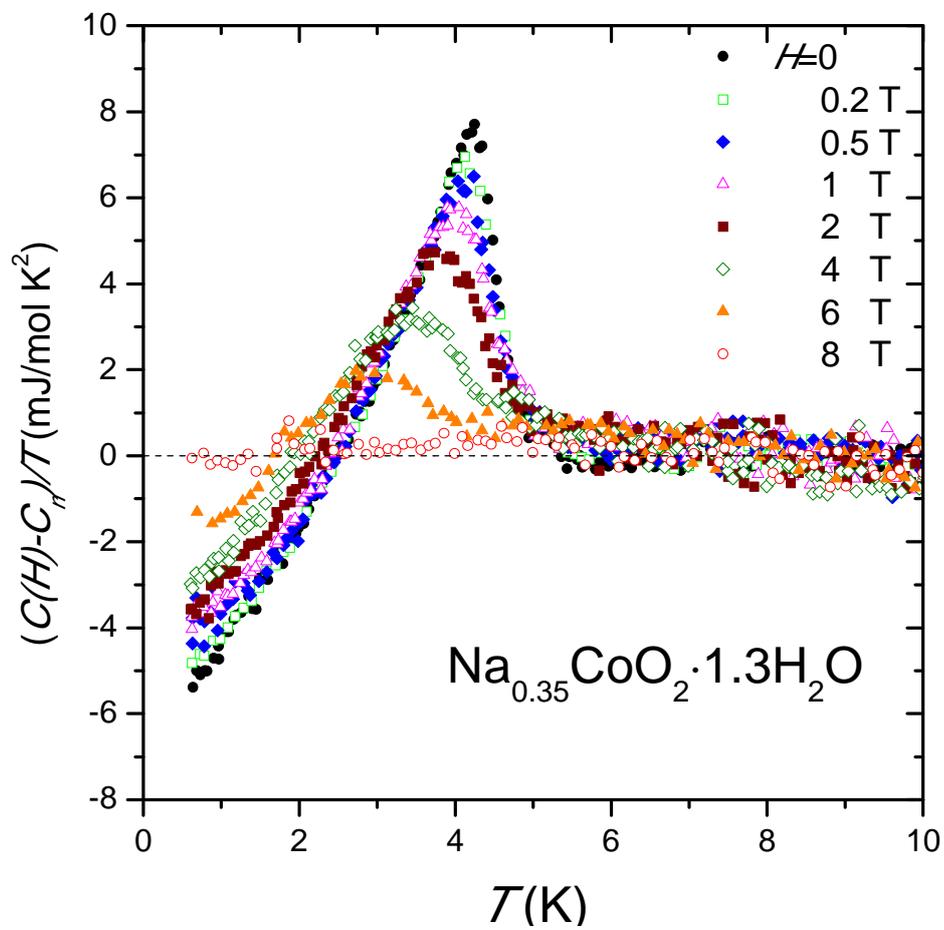

Fig. 4 Yang et al.

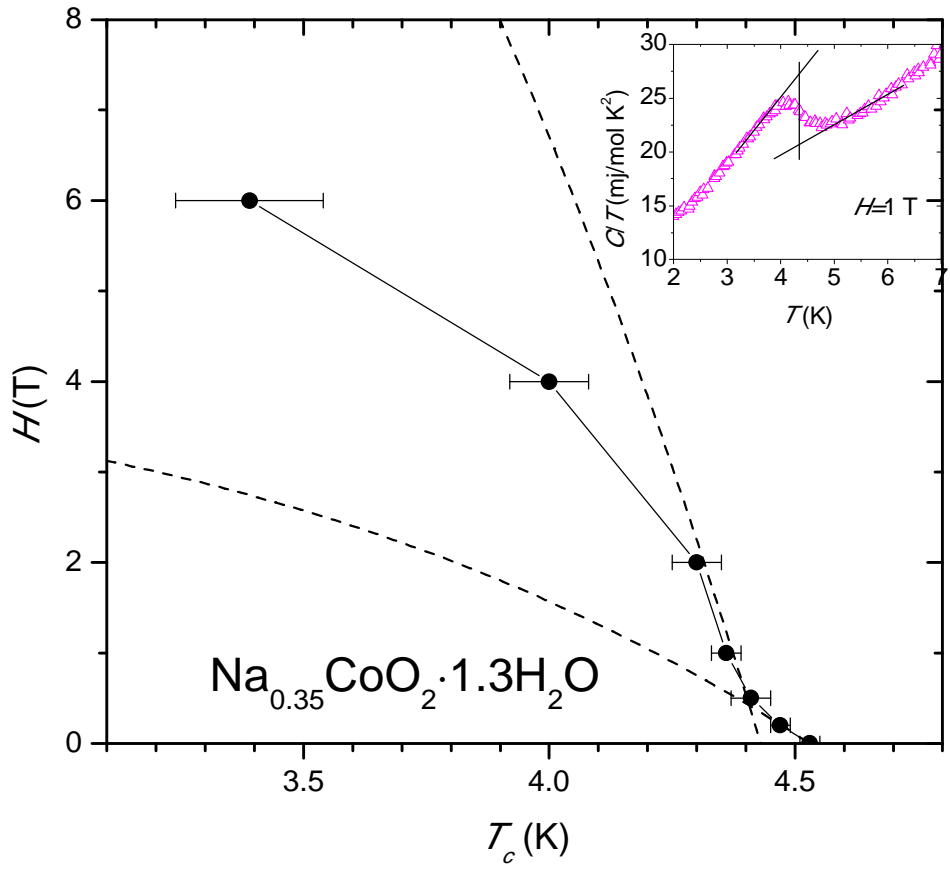

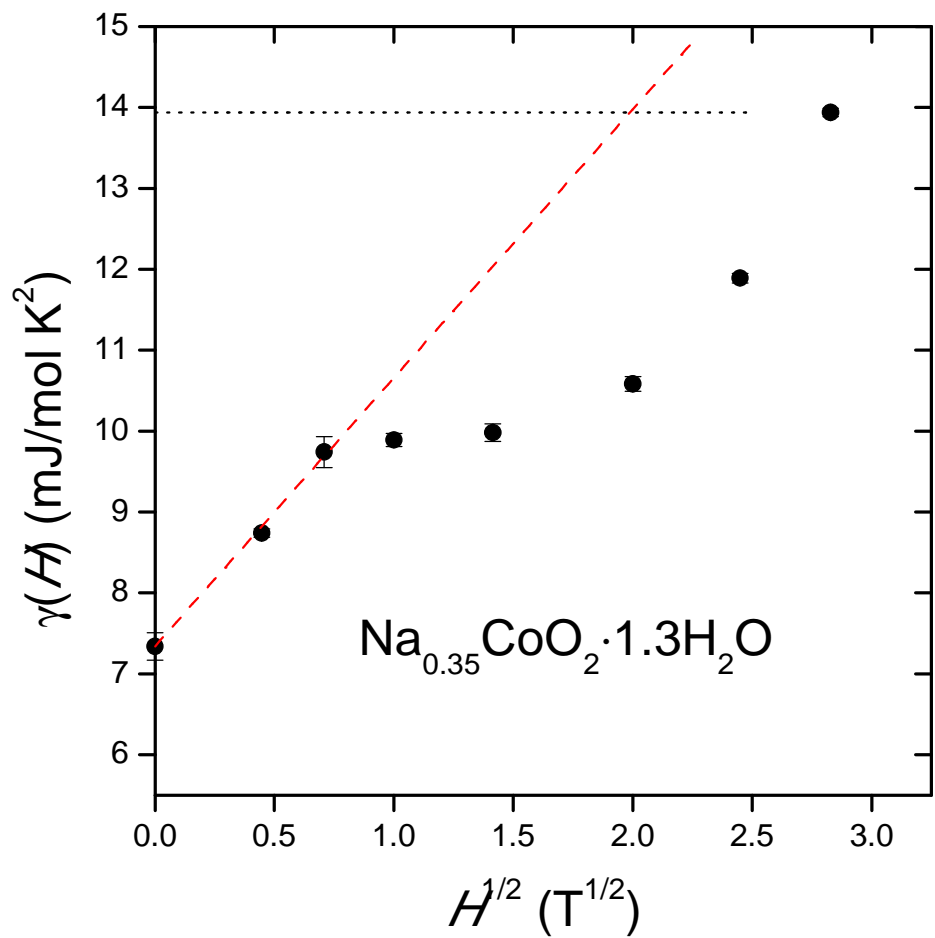

Fig. 5　Yang et al.